\documentclass[prb,twocolumn,amssymb,showpacs]{revtex4}
\usepackage{graphicx}
\usepackage{amsmath}

\begin{document}

\title{Investigation of the magnetic fluctuations in Tb$_2$Sn$_2$O$_7$ ordered spin ice by high resolution energy-resolved neutron scattering.}
\author{I. Mirebeau$^1$, H. Mutka$^2$, P. Bonville$^3$, A. Apetrei$^1$, A. Forget$^3$.
\\
}
\address{
$^1$Laboratoire L\'eon Brillouin, CEA-CNRS, CE-Saclay, 9191
Gif-sur-Yvette, France.}
\address{$^2$Institut La\"ue Langevin, 6 rue Jules Horowitz, BP 156X, 38042
Grenoble France.}
\address {$^3$Service de Physique de l'Etat Condens\'e,
CEA-CNRS, CE-Saclay,  91191 Gif-Sur-Yvette, France.}

\begin{abstract}
We have studied magnetically frustrated Tb$_2$Sn$_2$O$_7$ by
 neutron diffraction and high resolution energy-resolved neutron
 scattering. At 0.1 K, we observe short range magnetic
 correlations with a typical scale of 4 \AA,
 close to the near neighbor distance between Tb$^{3+}$ ions. 
  This short range order coexists with ferromagnetic correlations and long range spin
  ice order at the scales of 18 and 190 \AA, respectively.
 Spin dynamics was investigated at a time scale down to 10$^{-9}$s, by energy-resolved
 experiments on a backscattering spectrometer. We observe a freezing of the
 spin dynamics for all length-scales,
 with a strong slowing down of the spin fluctuations when long range order settles in.
 We discuss the spin fluctuations remaining in the ground state
 in comparison with previous data obtained by muon spectroscopy.

\end{abstract}

 \maketitle

Geometrical frustration is expected to favor the onset of unusual
types of order such as spin liquid and spin ices, showing a large
degeneracy of the magnetic ground state. In spin ices, it leads to
non zero ground state entropy akin to that of real ice
\cite{Ramirez99}, peculiar freezing dynamics\cite{Snyder01}, and
short range magnetic orders\cite{Bramwell01,Bramwell012}.
Application of a magnetic field lifts the ice rule degeneracy
\cite{Fukazawa02}, yielding to excitations which span the entire
system\cite{Jaubert08}, or are akin to magnetic
monopoles\cite{Castelnovo08}, depending on the orientation of the
field with respect to the anisotropy axis.

 Canonical spin ices are
 observed in pyrochlores R$_2$Ti$_2$O$_7$ (R=Dy or Ho),
 where the rare earth moments occupy a lattice of corner sharing tetrahedra.
 The stabilization of the spin ice ground state is related to the strong anisotropy
 of the 
  Dy$^{3+}$ or Ho$^{3+}$ ions, whose moments
 are constrained to lie along the $<$111$>$ local axes
 connecting the center of each tetrahedron to the summits.

The Tb pyrochlores offer a more complex but even richer behavior,
due to the smaller anisotropy of the Tb$^{3+}$ ion, and to the
fact that superexchange and dipolar interactions between
near-neighbor Tb$^{3+}$ ions nearly compensate. Tb$_2$Sn$_2$O$_7$
is an intriguing example of an ordered spin ice\cite{Mirebeau05}.
Contrarily to classical spin ices which do not order at large
scale, here the four tetrahedra of the unit cell are identical,
yielding magnetic Bragg peaks and long range order, at a length
scale which increases with decreasing temperature T and reaches
about 190 \AA\ at T=0. The onset of
 this magnetic order is observed at T$_{\rm I}$=1.3(1) K. An upturn of the
correlation length and magnetic moment, together with a peak in
the specific heat, occurs at T$_{\rm C}$=0.87(2)K.

Magnetic fluctuations play a prominent role in the ordered spin
state of Tb$_2$Sn$_2$O$_7$. They were first evidenced by the
reduction of the ordered moment as measured using the hyperfine
Schottky anomaly of the specific heat and compared with its value
from neutron diffraction\cite{Mirebeau05}. Then they were directly
observed using $\mu$SR \cite{Dalmas06,Bert06}. Surprisingly, no
evidence of any static component appears in the local field probed
by $\mu$SR in spite of the presence of the magnetic order. The
fluctuations of the local field occur at a time scale estimated to
either to  8 10$^{-11} s$ or to 5 10$^{-9}$ s, depending mostly on
the value assumed for the local field\cite{Dalmas06,Bert06}. Under
applied field, damped oscillations in the muon polarization, and a
thermal hysteresis in the longitudinal relaxation rate, show the
presence of local configurations frozen at the time scale of the
muon probe\cite{Bert06}.

Very recently, several inelastic neutron scattering measurements
were performed to study the magnetic fluctuations in
Tb$_2$Sn$_2$O$_7$.
 Crystal field
excitations\cite{Mirebeau07} were measured down to 1.4 K and compared to those in 
 Tb$_2$Ti$_2$O$_7$, which remains spin liquid
down to about 50 mK. It was shown that the wave functions
describing the ground and first excited states of the Tb$^{3+}$
ion are exchanged in the two compounds \cite{Mirebeau07}. However,
these differences in the local states of the Tb$^{3+}$ ion cannot
explain the different cooperative ground states in the Ti and Sn
compounds within the frame of current theories.

Neutron spin echo (NSE) measurements\cite{Chapuis07} were
performed in Tb$_2$Sn$_2$O$_7$ for a moment transfer $Q$=0.08
\AA$^{-1}$, outside the Bragg peak positions, probing the dynamics
of the ferromagnetic correlations. They clearly observed some spin
dynamics down to 0.8 K, but failed to detect any below. Finally,
the magnetic cross section was measured by polarized
neutrons\cite{Rule07}. The spin dynamics was deduced from the
comparison between diffraction measurements (integrating the
magnetic signal over all energies up to 3 meV) and elastic
scattering with a coarse energy resolution (integration over an
energy window of $\pm$0.15 meV). At 0.04 K, the authors conclude
to the presence of static Bragg peaks, together with a liquid-like
structure factor arising from spins moving faster than 0.04 THz.

Here we have investigated the spin correlations and dynamics in
Tb$_2$Sn$_2$O$_7$ by combining neutron diffraction and energy
resolved neutron scattering measurements, with a very high energy
resolution of 0.5 $\mu$eV at half width half maximum, probing time
scales up to 1.3 $10^{-9}$~s. Both measurements were performed in
the temperature range 0.1 K-1.4 K. The $Q$ and $T$ dependence of
the magnetic scattering was analyzed quantitatively, considering
not only the Bragg scattering, but also the medium and short range
correlations. Diffraction data measured at 0.1 K show the
coexistence of several length scales, corresponding to first
neighbors ($\sim$4 \AA), medium (18 \AA), and long range (190 \AA)
correlations. The spin dynamics was investigated in different $Q$
ranges where either short or long length scales are
 dominant. When $T$ decreases, we observe a freezing
of the spin dynamics for all length scales, with a stronger effect
in the region of the transition. Some spin fluctuations persist
down to the
lowest temperature of 0.1 K. 
We discuss these results with respect to the fluctuations in the
ground state in Tb$_2$Sn$_2$O$_7$.

\begin{figure} [h]
\includegraphics* [width=\columnwidth] {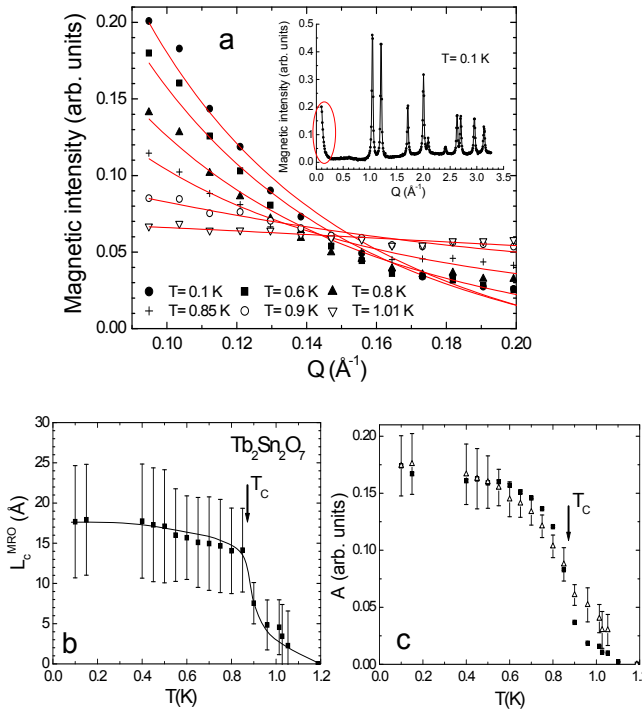}
\caption{(Colour online) a: Magnetic intensity measured
Tb$_2$Sn$_2$O$_7$ by neutron diffraction at small angles, versus
the moment transfer $Q$
for several temperatures. A spectrum above 1.2 K was subtracted.
Solid lines are Lorentzian fits (see text). In inset, the total
diffraction pattern at 0.1 K, showing the small angle intensity
together with the magnetic Bragg peaks. b: medium correlation
length L$_{\rm C}$ $^{\rm MRO}$ = 1$/$$\kappa$ (full squares)
versus temperature $T$. The solid line is a guide to the eye. c:
norm A of the Lorentzian (open triangles) and intensity of the
magnetic Bragg peaks (full squares) versus temperature. }
\label{fig1.eps}
\end{figure}
\section{Neutron diffraction}
Neutron diffraction measurements were performed on the
 D1B spectrometer at the Institut La\"ue Langevin (ILL) with an
incident neutron wavelength of 2.52 \AA.  The magnetic long range
order was already studied in ref. \onlinecite{Mirebeau05} from the
same data. The magnetic cross section was isolated by subtracting
a spectrum measured at 1.2 K. The magnetic diffraction pattern at
0.1 K (inset Fig 1a) shows an intense small angle neutron
scattering (SANS) signal for $Q$ $<$0.2 \AA$^{-1}$, arising from
ferromagnetic correlations. The SANS signal (Fig. 1a) strongly
increases with decreasing T, and flattens in the range of the
transition. It was fitted by a Lorentzian function
$I($Q$)=\frac{A}{\pi}\frac{\kappa}{\kappa ^2+ Q^2}$. Good fits are
obtained down to about 0.6 K, but the fit quality decreases below.
The fit however allows one to estimate the medium correlation
length L$_{\rm C}$ $^{\rm MRO}$ = 1$/$$\kappa$ (Fig. 1b) versus
temperature in the ordered region. With decreasing temperature,
L$_{\rm C}$ $^{\rm MRO}$ increases from values close to the near
neighbor distance  between Tb$^{3+}$ moments (d=3.686 \AA), to
about 18(6) \AA\ below 0.8 K. This value is about 10 times smaller
than that deduced from the width of the magnetic Bragg peaks
(L$_{\rm C}$=190 \AA\ as shown in ref. \onlinecite{Mirebeau05})
but its temperature dependence is quite similar. The parameter A
increases with decreasing temperature. Its variation compares well
with that of the LRO squared magnetic moment deduced from the
Bragg peak intensity, although it is less sharp in the transition
region (Fig. 1c).

\begin{figure} [h]
\includegraphics* [width=\columnwidth] {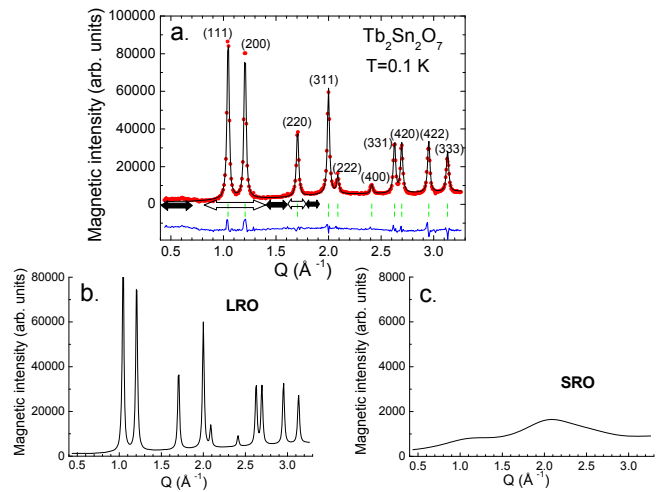}
\caption{(Colour online) a: Magnetic diffraction spectrum  of
Tb$_2$Sn$_2$O$_7$ versus the moment transfer
 $Q$ 
 at 0.1 K. A spectrum at 1.2 K was subtracted. The solid line is a refinement
 taking into account both LRO and SRO contributions.
 A linear background is also added. The arrows correspond to the Q-bands selected in the time-resolved experiment. White arrows correspond to the Bragg Q-bands,
  black arrows to the low, medium and high $Q$-bands.; b: (resp c:) calculated LRO (resp. SRO)
 contribution to the diffraction pattern. }
 \label{fig2.eps}
\end{figure}
\begin{figure} [h]
\includegraphics* [width=8 cm] {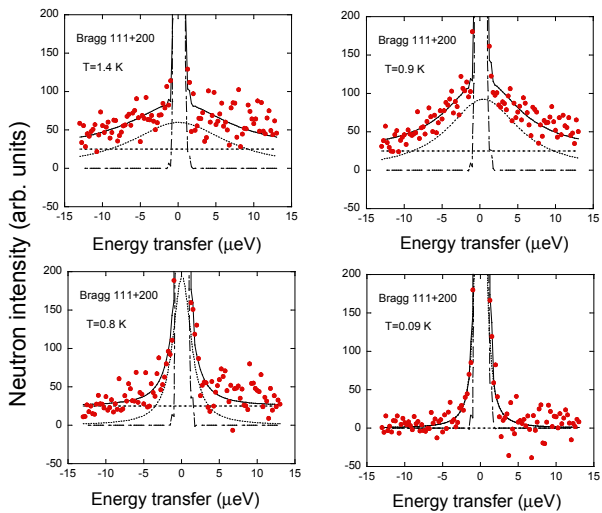}
\caption{(Colour online) Inelastic neutron spectra of
Tb$_2$Sn$_2$O$_7$ for several temperatures measured in the
$Q$-range of the 111 and 200 Bragg peaks. The solid line is a fit
as described in the text. Dotted, dashed-dotted and dashed lines
correspond to the quasielastic, elastic and flat $"$background$"$
terms respectively.} \label{fig3.eps}
\end{figure}

\begin{figure} [h]
\includegraphics* [width=0.8\columnwidth] {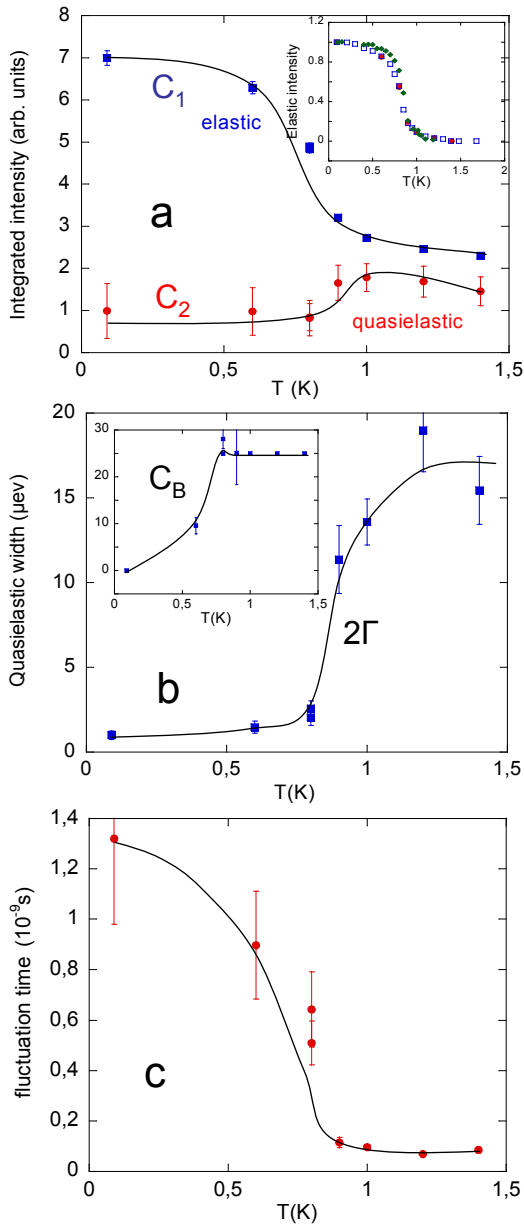}
\caption{(Colour online) Temperature dependence of the parameters
deduced from the
fit of the inelastic spectra 
in the region of the 111 and 200 Bragg peaks. a: elastic intensity
C$_1$ and quasielastic intensity C$_2$ deduced from the fit of the
energy patterns. In inset: the elastic intensity (full dots)
 is compared to the intensity from elastic scans (open squares)
 and to the integrated intensity
 of a magnetic Bragg peak measured in diffraction (diamonds).
 All quantities are scaled to vary between 1 at 0.1 K and 0 at 1.4 K.
b: quasielastic width 2$\Gamma$ versus temperature. In inset: the
background level C$_{\rm B}$, involving fluctuations at a shorter
time scale than probed in the energy window. c: the fluctuation
time
$\tau$=$\hbar$/$\Gamma$. 
Solid lines
are guides to the eye.} \label{fig4.eps}
\end{figure}
\begin{figure} [h]
\includegraphics* [width=0.8\columnwidth] {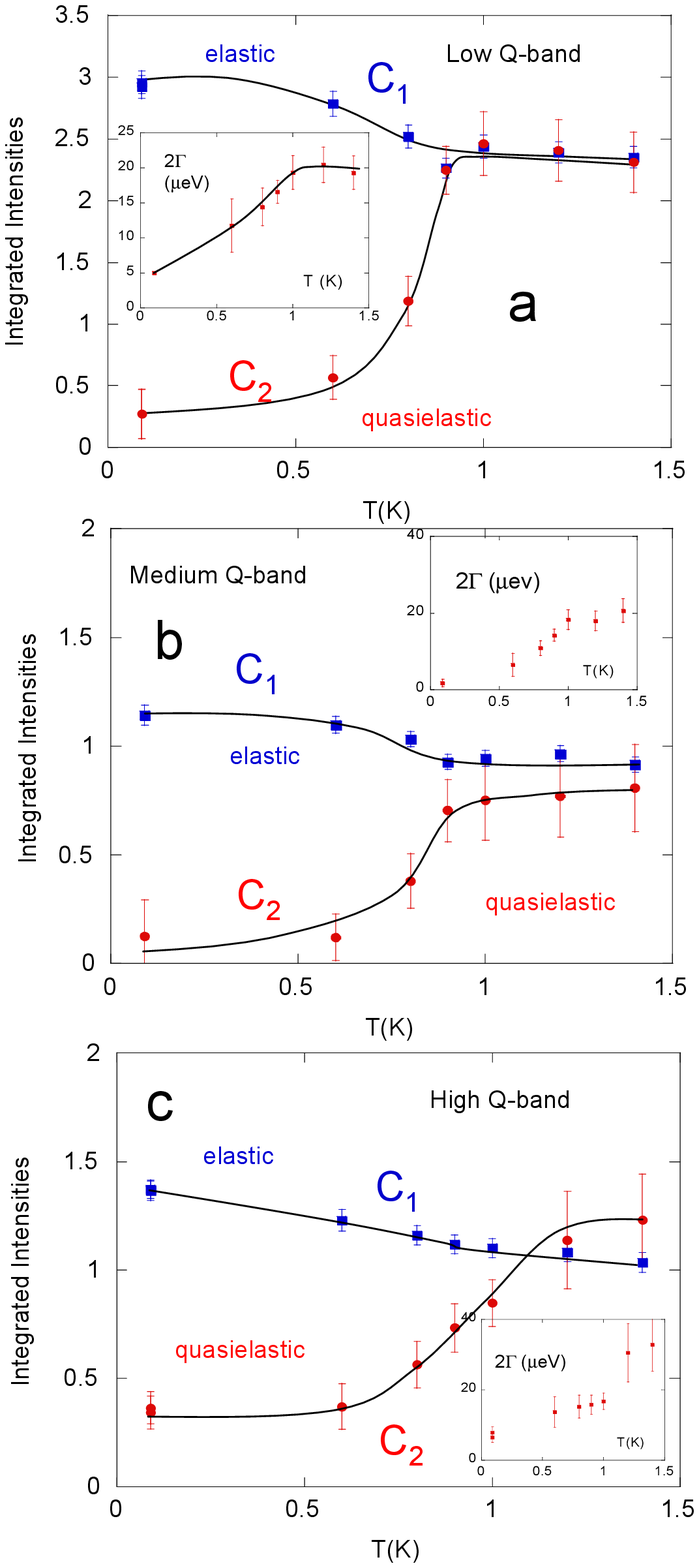}
\caption{(Colour online) Temperature dependence of the integrated
elastic and quasi elastic intensities.The background level was
fixed to zero at all temperatures. Solid lines are guides to the
eye. Fig. 5a, 5b, and 5c correspond to low, medium and high
$Q$-bands respectively. In the insets : the quasielastic width
2$\Gamma$ versus temperature. The width at 0.09 K in the inset of
Fig. 5a was fixed.} \label{fig5.eps}
\end{figure}
The short range order (SRO) yields a diffuse magnetic scattering
which remains clearly visible at 0.1 K. This scattering was not
analyzed in ref. \onlinecite{Mirebeau05}. Here we refined it as a
magnetic structure, assuming that it has the same symmetry as the
LRO, but a much shorter correlation length. The relative
contributions of the SRO and LRO are shown in Fig. 2. From the
fit, we obtain values of LRO and SRO moments of 5.8(1) and 3.3(3)
$\mu_{\rm B}$ respectively. The value of the LRO moment, ordered
at the scale of 190 \AA\ deduced from the width of the Bragg
peaks, agree with the value previously determined of
5.9(1)$\mu_{\rm B}$. The SRO moment compares well with the value
of about 3 $\mu_{\rm B}$ obtained by calibration of the diffuse
magnetic scattering measured with polarized neutrons\cite{Rule07}.
This analysis shows that a large amount of the Tb$^{3+}$ moments
remains ordered at a very short length scale, estimated here to
4(1) \AA\ from the refinement, therefore comparable to the
distance between first neighbor Tb$^{3+}$ moments (3.686 \AA).
Assuming that both types of correlations contribute to the neutron
intensity independently, we estimate the total ordered moment to
M= 6.6(2) $\mu_{\rm B}$ $/$Tb. We notice that this value is
exactly the value calculated for the total local moment within the
crystal field model (Fig. 15 of ref. \onlinecite {Mirebeau07}).

The above analysis shows the presence of three characteristic
length scales for the spin correlations at low temperature, with
values ranging from the first neighbor distance to length scales
of about
 18 cubic unit cells (the lattice constant a = 10.426 \AA).
 A distribution of length scales was previously inferred
from an independent analysis of the profile peak
shape\cite{Chapuis07}.

\section{Energy resolved neutron scattering}
 The  analysis above motivates  a  study  of  the slow  spin
dynamics  at different $Q$-ranges, where either short or long
length scales dominate. For this purpose we performed energy
resolved neutron scattering measurements on the IN16
backscattering spectrometer of the ILL, with an incident neutron
wavelength of 6.27 \AA. This instrument  offers  a dynamic range
of  $\pm$ 15 $\mu$eV for the observation of fluctuations
characterized by a quasielastic response. Due to the large angular
acceptances of the backscattering method the wave-vector transfer
resolution is rather relaxed, anyhow sufficient to discriminate
regions of interest with or without contributions  from the Bragg
peaks associated with the long-range ordering process. Optionally
one can switch off the  energy transfer analysis  to follow only
the elastic response  within the resolution window of 1 $\mu$eV at
full width half maximum (FWHM). Accordingly one can obtain a
temperature scan of the elastic contribution over the whole
$Q$-range and compare intensity in specific $Q$-bands  to the
temperature variation observed
 for the Bragg peaks in the diffraction experiment.

 The sample was packed in a 1 mm thick flat can in order to
minimize the sample absorption. The transmission was evaluated to
0.9. The counting time was 12 hours for each temperature. A
vanadium sample of the same shape
 was measured to calibrate the angle dependent detection
 efficiency and determine the energy resolution.

 The spectra were corrected for background
and absorption using spectra of the empty sample holder measured
in the same conditions, taking into account the specific
corrections due to the backscattering geometry \cite{Randl96}.
Temperature scan of the elastic resolved scattering was taken  in
the range 0.1$\leq T<$1.4 K. The quasielastic spectra were
recorded in a range 0.1$<Q<$1.9~\AA$^{-1}$. They were grouped into
five $Q$-bands, three bands out of the Bragg peaks, and two at the
Bragg peak positions. The $Q$-bands outside the Bragg peaks
 are respectively called the low $Q$-band,
 ($Q= 0.55\pm0.15~$ \AA$^{-1}$), the medium $Q$-band ($Q= 1.5\pm0.1~$ \AA$^{-1}$)
 and the high $Q$-band ($Q= 1.85\pm0.05~$\AA$^{-1}$).
 The Q-bands at the Bragg peak positions are respectively
  $Q_{111+200}= 1.1\pm0.3~$ \AA$^{-1}$ and  $Q_{220}= 1.7\pm0.1~$
  \AA$^{-1}$,
the first for the 111 and 200 together (not well resolved
individually) and the second for 220 contribution. The 311 Bragg
peak is out of the accessible $Q$-range. We have situated these
$Q$-bands in the diffraction pattern of the Fig. 2. One sees that
the Bragg $Q$-bands  mostly probe the long length scales (the
Bragg intensity dominates), whereas the low, medium and high-$Q$
bands probe the
 shortest lengthscale. 
 We also
note that the nuclear Bragg contribution on all of the above peaks
is either zero or negligible, and that the overall nuclear
incoherent
  contribution is also small due to the small incoherent cross-sections of the elements.
By fitting the vanadium patterns
  with a Gaussian function, we get
an energy resolution at half width half maximum (HWHM) of 0.4
$\mu$eV in the Bragg peak and large angle regions, and 0.55
$\mu$eV in the low angle region.

 Typical energy resolved neutron scattering patterns are shown
in Fig. 3, in the $Q$-band covering  the 111 and 200 Bragg peaks.
They clearly show a quasi-elastic signal in the temperature range
0.8-1.2 K. The quasielastic signal strongly narrows with
decreasing temperature, which reflects the freezing of the spin
dynamics around the transition. The neutron intensity was fitted
by the sum of three contributions. First, an elastic (or
resolution-limited) signal of intensity C$_1$, of magnetic origin
mostly, accounts for the spins which fluctuate at time scales
longer than the resolution limit.
 Second, a Lorenzian function of intensity C$_2$ and half-width at half
maximum (HWHM) $\Gamma$ accounts for the quasi elastic signal. The
Lorenzian quasi elastic signal was convoluted by the instrumental
resolution. An energy-independent flat level $"$background$"$
C$_{\rm B}$ accounts for the $"$fast$"$ fluctuations (or short
time scales), corresponding to energy widths larger than the
energy window. Above 1 K, the flat level and the rather broad
Lorentzian contributions cannot be determined independently with
good accuracy. So, the value of C$_{\rm B}$ was fixed to the value
at 0.9 K. Below 0.6 K, the quasielastic linewidth has the same
order of magnitude as the resolution, and the flat level vanishes.

The parameters deduced from the fits are plotted versus
temperature in Fig. 4 in the $Q$-band of the 111 and 200 Bragg
peaks. Similar results are obtained for the $Q$-band  of the 220
peak.
 The elastic intensity C$_1$ increases with
decreasing $T$. Its temperature variation scales quite well with
that of the Bragg intensity deduced from diffraction data. The
latter represents a constant angle integral  of the scattering
cross-section over energy transfers
 up to the  incident neutron
energy (inset Fig. 4). Just below T$_{\rm C}$=0.87(2)K, the
intensity measured by diffraction increases more sharply with
decreasing $T$ than the elastic intensity. The differences in the
$T$-dependence of the Bragg peak and the elastic resolved
intensities can be explained by two factors, one being the
possibility of integrating the $"$fast$"$ dynamics (represented by
the flat level in the energy resolved experiment), the other the
coarser $Q$-resolution of the time resolved patterns, which
integrate a contribution from the SRO not included in the Bragg
peaks of the diffraction patterns.

The quasielastic intensity C$_2$ shows a broad maximum in the
transition region, where the contribution from the fast
fluctuations starts to enter the energy window. As $T$ decreases
further, C$_2$ starts to decrease when the spectral weight is
transferred to the elastic intensity. The energy width 2$\Gamma$
decreases rapidly  down to 0.8 K; a quasielastic signal of width
close to the resolution limit remains at 0.09 K. The energy
independent flat level C$_{\rm B}$ also decreases with decreasing
temperature. At 0.09 K, there is no visible contribution of the
 $"$fast$"$ fluctuations to the energy pattern and C$_{\rm B}$ was fixed
to zero.

 The
temperature variations of the parameters C$_1$, C$_2$, C$_{\rm B}$
and $\Gamma$ reflect a freezing of the spin dynamics. The elastic
contribution increases at the expense of the $"$fast$"$
fluctuations and quasielastic contributions, whereas the decrease
of the linewidth reflects the increase of the relaxation time
probed in the time window of the measurement.

The Lorentzian line shape of the quasielastic signal corresponds
to a relaxation process which involves spin fluctuations with a
lifetime $\tau$ ($<$S(0).S(t)$>$$\propto$ e$^{-t/\tau}$). Taking
the Fourier transform of this expression yields
$\tau$=$\hbar$$/$$\Gamma$, where $\hbar$=0.658 10$^{-9}$ $\mu$eV.
s (see note \onlinecite{note2pi}). 
The $T$ dependence of the
relaxation time is shown in Fig. 4 c. At 0.09 K,
 the time scale of the spin fluctuations is equal to $\tau$$_{0}$=1.3
10$^{-9}$~s ($\Gamma$= 0.5 $\mu$eV).

Energy spectra in the low, medium and high $Q$-bands also probe a
freezing of the spin dynamics. Here, no contribution of a flat
level was detected and the fits were performed with C$_1$ and
C$_2$ contributions only. Their variations (Fig. 5) reflect those
at the magnetic Bragg peaks. The $\Gamma$ values at 0.09 K are
higher, probing fluctuations at shorter time scales ($\tau$$_{0}$
$\sim$ 0.3-1 10$^{-9}$~s). The relative weight of the
$"$fluctuating$"$ spins with respect to the $"$static$"$ spins at
low temperature is also higher in the $Q$-bands outside the Bragg
peaks. Here the terms $"$static$"$ and $"$fluctuating$"$ refer to
the time scale defined by the instrumental resolution of IN16 (1.3
$10^{-9}$~s) and sensitivity of the fit.

One notices that due to the limited time window and the strong
change of the correlations with temperature, the opposite
variations of the quasielastic and elastic intensities during the
freezing process do not compensate in a given $Q$-band. Namely, at
the Bragg peaks the $"$loss$"$ in the quasielastic intensity with
decreasing temperature is much smaller than the $"$gain$"$ in the
elastic intensity, whereas the reverse situation
occurs outside the Bragg peaks. 
\section{Discussion}
The above features show a distribution of time scales, which
extends to longer times with decreasing temperature, as the
characteristic length scales of the spin correlations increase.
Somewhat similar effects are seen in other frustrated magnets. A
distribution of fluctuation times concomitant with a distribution
of correlation lengths is observed by neutron scattering in
disordered systems such as spin glasses \cite{Murani,Bellouard97}.
 A broad range of time scales in absence of
long-range order but  with remarkably energy(time)-scale invariant
short-ranged correlations was also seen in the well-known
geometrically frustrated Kagome spin
 system SCGO
\cite{Mutka}.

 The time scale of the fluctuations is in rather good agreement with the time scale probed by neutron spin
 echo (NSE) experiment\cite{Chapuis07} in the range of the transition ($\tau$
$\sim$ 0.2 10$^{-9}$~s for T=0.8 K
 and $Q$=0.08 \AA$^{-1}$). At 0.1 K, when NSE did not
 observe further dynamics, our experiment in the backscattering geometry
 still suggests fluctuations at a time scale
 $\tau$$_{0}$ with $\tau$$_{0}$ ranging from 1.3 to 0.3
 10$^{-9}$s, the shortest time scales corresponding to the SRO.
Fluctuations at longer time scales than $\tau$$_{0}$ do exist (the
elastic scattering),
  but no faster fluctuations are probed
  (there is no residual background in the energy window).

  The life time of the spin fluctuations may be compared to
that probed by $\mu$SR experiments\cite{Dalmas06,Bert06}. In ref.
\onlinecite{Bert06}, the $\mu$SR response was interpreted by a
relaxation process between different configurations of the local
field B$_{loc}$. The typical time $\tau_c$ of this process
  can be evaluated from the longitudinal relaxation rate
  $\lambda_Z$ by the relation
  $\tau_c$$\sim$ $\lambda_Z$$/$($\gamma$$_{\mu}$B$_{loc})^2$ where $\gamma$$_{\mu}$
  is the muon gyromagnetic ratio. At 0.1 K $\tau_c$ could be evaluated to 8
  10$^{-11}$ s in ref. \onlinecite{Dalmas06} (taking B$_{loc}$=0.2 T) and 5 10$^{-9}$ s in ref.
\onlinecite{Bert06} (taking B$_{loc}$=0.02 T). Our estimation of
the life time ($\tau$$_{0}$ $\sim$0.3-1.3
 10$^{-9}$s) situates in between these values.

 On the other hand, the
fluctuations at longer time scales than $\tau$$_{0}$, clearly seen
by neutrons (they give rise to the elastic scattering and to a
static response in the NSE experiment), do not yield any static
component in the muon spectra. A strong fully static component
cannot be excluded by the neutron data, but seems to be
incompatible with the muon ones. To investigate this point
further,
we computed the muon depolarization in presence of a dipolar field
randomly reversing with the frequency $\nu$, as done in ref.
\onlinecite{Bonville08}. This shows that fluctuations with a
frequency $\nu$$>$2$\gamma$$_{\mu}$B$_{loc}$
 should wash out the precession signal, clear-cut
signature of a static field in the muon data. For
B$_{loc}$=0.02~T, this yields a higher limit of 2 10$^{-8}$s for
the typical time scale $\tau$=1/$\nu$ of the
fluctuations\cite{note2pi}, which is still compatible with the
lower limit of 1.3 10$^{-9}$s given by the instrumental resolution
of IN16. This is no longer true for the higher value
B$_{loc}$=0.2~T. Comparison between muon and neutron data
therefore favor a lower estimation of the internal field.

The fluctuations of the local field in ordered spin ice
Tb$_2$Sn$_2$O$_7$ found by $\mu$SR found by are quite similar to
those observed in the parent spin
 liquid Tb$_2$Ti$_2$O$_7$. In both compounds, the short scale fluctuations
may involve tunnelling or thermally activated excitations between
the six degenerated configurations of the spin ice structure. In
ordered Tb$_2$Sn$_2$O$_7$, such excitations may also help to
$"$switch$"$ domains of longer length scales.

In Tb$_2$Sn$_2$O$_7$, our results suggest a homogeneous picture
where different correlation lengths involve different spin
components and different time scales, rather than a coexistence of
independent clusters with different sizes.
 In the pyrochlore
spinel ZnCr$_2$O$_4$ and related compounds, a cluster-like
scattering process has been invoked, with the presence of emergent
excitations within loops of strongly bound hexagonal clusters
\cite{Lee02}. More recently however, theoretical work on the
$"$true$"$ spin ice Dy$_2$Ti$_2$O$_7$ showed that this picture
does not necessarily imply real independent clusters and that long
range interactions are relevant\cite{Yavors08}. Whether such
picture can be extended to describe excitations in the ground
state of the ordered spin ice Tb$_2$Sn$_2$O$_7$, in which we
observed multiple length scales, remains a matter for further
theories.

\section{Conclusion}
In conclusion, in Tb$_2$Sn$_2$O$_7$ ordered spin ice we have
observed a freezing of the spin fluctuations which involves the
coexistence of different correlations lengths and times scales.
The distribution extends to longer times with decreasing
temperature. A marked slowing down is observed in the transition
region, in agreement with previous muon and neutron spin echo
experiments. At the lowest temperature, the time scale of the spin
fluctuations ranges between 1.3 and 0.3 10$^{-9}$ s.
 Fluctuations at longer time scales are also observed by neutrons,
 whereas they give no clear-cut signature in the muon spectra.
 This may be compatible with the low estimation of the local
 field (B$_{loc}$=0.02~T) found in $\mu$SR. The presence of
 multiple time scales remains to be understood in further
 theoretical work.


We thank B. Frick and R. Chung  for the  help with the back
scattering neutron experiment, O. Isnard for the help with the
neutron diffraction experiment, and J.-L. Ragazzoni
 for setting up the dilution equipment.


\begin{thebibliography}{}
\bibitem{note2pi}The equation which yields the typical time scale holds
because it involves an $\it{energy}$ measurement and a Fourier
transform. This is the case for Quasielastic or Inelastic Neutron
Scattering, Electronic Paramagnetic Resonance, M\"ossbauer
spectroscopy or energy-resolved Nuclear Magnetic Resonance (NMR).
The time involved results from a dynamical broadening or from the
peak lineshape associated with the physical process, so that the
characteristic time is the inverse of the pulsation $\omega$,
namely $\tau$= 1$/$$\omega$=1$/$$2\pi\nu$, where $\nu$ is the
frequency equivalent to the energy $\hbar\omega$ involved in the
process. On the reverse, when the typical time scale is determined
from a $\it{time}$ measurement such as muon spectroscopy or
time-resolved NMR, where the process involved is expressed as
e$^{-t/\tau}$, then $\tau$=1$/$$\nu$ is expressed as the inverse
of the frequency. This is also the case when $\tau$ is simply
$\it{defined}$ with respect to the frequency of a physical event,
as being the average time between two events.
\end{thebibliography}
\end{document}